\documentstyle[sprocl,amstex,epsfig]{article}

\def\be{\begin{equation}}
\def\ee{\end{equation}}
\def\bea{\begin{eqnarray}}
\def\eea{\end{eqnarray}}
\newcommand{\cor}[1]{\left\langle{#1}\right\rangle}
\newcommand{\xpr}{x_\perp}
\newcommand{\rr}[4]{#1, {\it #2 \/}{\bf #3} #4}

\begin{document}

\title{HIGH ENERGY SCATTERING FROM THE $AdS/CFT$ CORRESPONDENCE\footnote{Talk 
presented at the DIS00 workshop, Liverpool, April 2000.}
}

\author{R. PESCHANSKI}

\address{Service de Physique Th\'eorique, CEA, CE-Saclay\\
F-91191 Gif-sur-Yvette Cedex, France}


\maketitle\abstracts{We apply the AdS/CFT correspondence to derive expressions 
for the scattering amplitudes at high energy for gauge theories at strong 
coupling. A method  is proposed based on the computation of 
correlators of Wilson loop operators  by their stringy duals in AdS spaces using 
either the supergravity (weak field) or classical (minimal surface) 
approximations.}

\section{Introduction}

Historically, the interpretation of strong interactions in terms of a string 
theory has raised much hope \cite{frampton} but was a deceiving adventure. 
Indeed, while the Veneziano (resp. Shapiro-Virasoro) amplitudes for Reggeon 
(resp. Pomeron) exchanges  were very promising  and at 
 the root of the open (resp. closed) string theories, problems 
arise when looking for the internal consistency of the whole scheme in our
4-dimensional world: a quantum anomaly requires 26 or 10 dimensions, gravitons 
and zero-mass vectors unavoidably appear in the spectrum of strong interaction 
states. So  a stringy description remains an open problem for  $QCD_4.$

Recently, the proposal of an AdS/CFT correspondence \cite{adscft} 
seems to overcome some difficulties met during the last 30 
years. In very brief terms (for a 
extended review, see  \cite{review}) the idea is to unify a ``microscopic''
and a  ``microscopic'' description of a
configuration of  $N \gg 1$ three-branes  in the so-called Type II-B string 
theory in 10 dimensions. 

In the ``microscopic'' description, the system gives rise to a 4-dimensional 
$SU(N)$ gauge theory (with $ {\cal N} = 4$ supersymmetries), while in the 
``macroscopic'' one, it is the source of  a gravitational background equipped 
with a $AdS_5  \otimes S_5$ metric with the physical Minkowski space lying at 
the boundary of $AdS_5 $. A duality property is 
conjectured between the  4-dimensional $SU(N)$ gauge theory at {\it strong} 
coupling and the gravitational background at {\it weak} coupling. Interestingly 
enough, the dynamical r\^ole of the  fifth dimension in  $AdS_5 $ is crucial for 
the validity of the correspondence.

The case of a gauge theory with $ {\cal N} = 4$ supersymmetries corresponds to a 
4-dimensional, non-confining, conformal field theory. The conjecture could be 
enlarged to confining theories without supersymmetry (e.g. see  \cite{witten}) 
by 
introducing a ``horizon'' scale in the 5-th dimension. The term ``horizon'' 
comes from the consideration of a black hole metric in the bulk of $AdS$ space 
in order to break supersymmetry. Even if the exact dual of $QCD_4$ has not yet 
been identified, these dualities give a laboratory framework for gauge 
observables at  strong coupling. For instance, the Wilson area criterion for 
confinement can be explicitely verified \cite{Wilson}.

\section{High energy amplitudes: obervables and results}

Let us briefly outline the derivation of our papers  \cite {janik1,janik2}. 
Scattering amplitudes in the high energy limit (and small
momentum transfer) can be
conveniently expressed in terms of a
correlator of Wilson  loops \cite{Nachtr}.
\begin{equation} 
\label{e.ampinit}
A(s,q^2) = -2is \int d^2\xpr e^{iq\xpr} 
\cor{\frac{W_1W_2}{\cor{W_1}\cor{W_2}}-1}
\end{equation}
where the two {\it tilted} Wilson loops  follow 
elongated trajectories  along the light-cone direction with transverse 
separation $a$ and a  tilting  angle $\theta$ around the impact parameter axis.
This corresponds to the scattering of colorless 
quark-antiquark pairs of mass $m \sim a^{-1}$. Indeed, the geometrical 
parameters of the configuration can be related to the
energy scales by analytic continuation
$\cos\theta \to \cosh \chi\equiv \frac{1}{\sqrt{1-v^2}}=\frac{s}{2m^2}-1$
where $\chi=\frac{1}{2}\log \frac{1+v}{1-v}$ is the Minkowski angle
(rapidity) between the two lines, and $v$ is the relative velocity.

The results we obtain distinguish between the large and the small impact 
parameter kinematics. At large impact parameter, we could use  \cite{janik1} the 
supergravity approximation of the appropriate type II-B string theory, since the 
fields, in particular gravity, are weak. We computed  the exchange contribution 
of all zero-mode fields between the two separated $AdS_5$ surfaces whose 
geometry is fixed by area minimization with the two initial
 Wilson loops at the boundary. Looking to 
the contribution  of the various fields (dilaton, Kaluza-Klein scalars, 
antisymmetric tensors mixed with Ramond-Ramond forms and the graviton) we find a 
hierarchy of real phase-shifts $\delta(b) \equiv \log 
\cor{\frac{W_1W_2}{\cor{W_1}\cor{W_2}}}$ contributing to elastic scattering at 
large impact parameter only. Indeed, this hierarchy is different from the 
static Wilson loop correlator  \cite {Wilson}, since the graviton is dominant 
and not the 
Kaluza-Klein scalars. The potential problems with unitarity are avoided, since 
the weak field approximation appears to be valid only at very large impact 
parameter $L/a \gg s^{2/7}$ where the scattering amplitudes are purely elastic. 
Note, however, the retentivity of the gravitational interaction, which  is still 
mysterious in the general  context of the AdS/CFT correspondence where the 
decoupling from gravity is expected.

In a second paper   \cite{janik2}, we addressed the problem of small impact 
parameter and the origin of inelasticity, i.e. imaginary contributions to the 
phase shift.  We concentrated on a situation where the difficulty with
supergravity field exchanges does not arise, since there exists a
single connected minimal surface which gives the dominant contribution
to the scattering amplitude in the strong coupling regime. This  allows us to 
extend our study to small
impact parameters, where inelastic channels are expected to play an
important r\^{o}le.
Moreover, it is possible to investigate both cases of 
conformal (non-confining) and confining cases by considering the appropriate 
geometries in $AdS$ spaces. Our goal was to understand the r\^ole of confining 
geometries in the characteristic features of scattering amplitudes at high 
energy. The main expected feature is  Reggeization, i.e. the 
determination of the amplitudes by singularities (poles and cuts) in the complex 
plane of the crossed channel partial waves, moving with $t \equiv q^2.$

Our main result  \cite{janik2} is that high energy amplitudes are governed by 
the geometry of minimal surfaces, generalizing  the {\it helicoid}  in 
 different $AdS$ geometries with the elongated  tilted Wilson loops at the 
boundary. Indeed, the confining geometries have the 
remarkable properties  to admit approximately flat configurations near the 
horizon scale in the fifth dimension and thus the tilting angle induces 
(approximate) helicoidal solutions for the minimal surface problem. For this 
solution and after analytic continuation, one finds a Regge singularity 
corresponding to a  linear double Regge pole 
trajectory with intercept one
\begin{equation}
\label{e.bhregge}
\alpha(t)= 1+\frac{R_0^2}{4 \sqrt{2g^2_{YM}N}}t\  ,
\end{equation}
where $R_0$ is the horizon scale and $g_{YM},$ the gauge theory coupling.

The results in confining geometries for impact parameter larger or of the order 
of $R_0$ can be contrasted with the conformal 
(non-confining) $AdS_5  \otimes S_5$ case which, using an asymptotic evaluation   
(the  mathematical knowledge on minimal surfaces embedded in $AdS$ spaces is yet 
limited!),
leads to  amplitudes with flat trajectories of the type 
\begin{equation}
\label{ampli}
A(s,t)\sim i s^{1+\frac{ 2\pi^4}{\Gamma(1/4)^4} \cdot
\frac{\sqrt{2g^2_{YM}N}}{2\pi}} 
t^{-1-\frac{F(\pi/2)}{2\pi} \ 
\frac{\sqrt{2g^2_{YM}N}}{2\pi}},
\end{equation}
where $F(\pi/2)\sim .3\pi$ comes from an anomalous dimension computed in 
 \cite{drukker}.

However, even in the confining cases, it may of course happen that the impact 
parameter distance between the
two Wilson  
lines becomes much smaller than $R_0.$ In this case (see Fig.~1)
the minimal surface problem becomes less affected by the black hole
geometry  and will just probe the small $z$ region of the
geometry.
 
The precise behaviour at these shorter distances will depend on the
type of gauge theory and, in particular, on the small $z$ limit of the
appropriate metric. In fact
 this limit resembles the original $AdS_5\times S^5$
geometry. Note  that the
same behaviour can be equivalently obtained through rescaling, by
keeping the impact parameter fixed and putting the scale $R_0 \to  \infty$. 
The conformal behaviour of the amplitude
(\ref{ampli}) may thus give a hint of 
the  small impact parameter limit also present in 
the physical confining case of $QCD_4,$ and thus a kind of hard-soft transition 
in impact parameter.
\begin{figure}[t]
\begin{center}
\mbox{\epsfig{figure=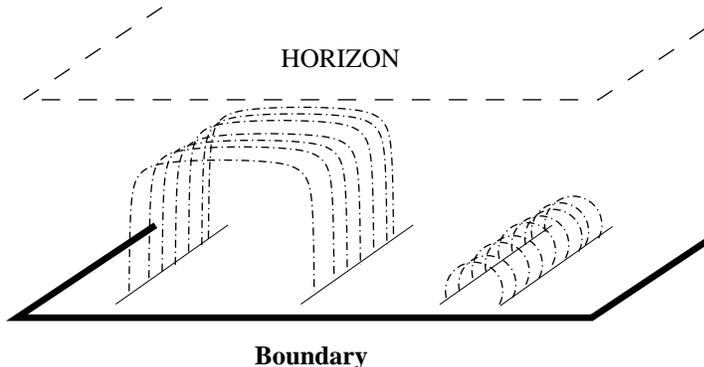,height=5.0cm}}
\vspace{-0.5cm}
\caption{The minimal surface in the black hole geometry. For simplicity the 
Wilson
lines are drawn here with vanishing angle of tilt $\theta=0$.}
\vspace{-0.5cm}
\label{fig:fig1}
\end{center}
\end{figure}
\section{Outlook}
Using the $AdS/CFT$ correspondence, we found  a relation between high-energy 
amplitudes in gauge theories at strong coupling and minimal surfaces 
generalizing the helicoid in various $AdS$ backgrounds.
We considered three
cases: (i) flat metric approximation of an $AdS$ black hole metric giving
rise to Regge amplitudes with linear trajectories, (ii) an
approximate evaluation for the conformal $AdS_5 \times S^5$ geometry
leading to flat Regge trajectories
and (iii) evidence for a transition, in a confining theory, from
behaviour of type (i) to (ii)
when the impact parameter decreases below the interpolation scale set by
the horizon radius. It would be quite useful to supplement the approximations 
made in our investigations by an evaluation of the string fluctuation pattern 
around the classical configurations we analyzed, in order to have a more precise 
determination of the predictions based on  the $AdS/CFT$ correspondence. After 
that, we will be able to discuss the validity and usefulness of this stimulating 
conjecture in a deeper way. 

\section*{Acknowledgments}

This work was done in tight collaboration with R. Janik whose name should be 
associated to all the results mentioned in this contribution. This work was 
supported in part by the EU Fourth Framework Programme `Training 
and Mobility of Researchers', Network `Quantum Chromodynamics and the Deep 
Structure 
of Elementary Particles', contract FMRX-CT98-0194 (DG 12 - MIHT).

\end{document}